# Magnetic Reversal and Critical Current Transparency of CoFeB Superconductor-Ferromagnet-Superconductor Heterostructures

Melissa G. Loving, Thomas F. Ambrose, Shawn Keebaugh, Donald L. Miller, Robert Pownall, Nicholas D. Rizzo, Anton N. Sidorov, Nathan P. Siwak

Advanced Technology Laboratory, Northrop Grumman Corporation, 1212 Winterson Road, Linthicum, MD 21090

In this work, we show fundamental low temperature ($T$) magnetic and $I_c$ responses of a magnetic Josephson Junction (MJJ) S/F/S heterostructure - Nb/ $Co_{56}Fe_{24}B_{20}$ /Nb. The ultra-thin $Co_{56}Fe_{24}B_{20}$ (CFB) films (0.6-1.3 nm) were deposited onto two separate buffer layers: 150 nm Nb/5 nm Cu and 150 nm Nb/ (1 nm Cu/0.5 nm Nb)$_6$/1 nm Cu. Both film sets were capped with 5 nm Cu/500 nm Nb. Magnetic results show reduced switching distributions in patterned arrays measured at near liquid Helium temperature (~ 10 K), with the incorporation of the (1 nm Cu/0.5 nm Nb)$_6$/1 nm multilayer. In electrical devices, the critical current ($I_c$) through the CFB layer decays exponentially with increasing ferromagnetic layer thickness and shows a dip in $I_c$ at 0.8 nm, characteristic of a change in the equilibrium Josephson phase in an S/F/S structure.

## I. Introduction

The rising demand for high efficiency computing has driven development pursuits to identify a "beyond-CMOS" dimensionally scalable logic technology. Superconducting digital technologies, such as reciprocal quantum logic (RQL) [1] and efficient rapid single flux quantum (eRSFQ) logic [2] offer a significant improvement in energy efficiency over CMOS technology, and are therefore under consideration for power-efficient, high-performance computing systems. [3] [4] Ultimately, such computing systems require a compatible low temperature memory solution that is power efficient, dense, and fast. To meet this need, Josephson magnetic random access memory (JMRAM) is being pursued. [5] [6] [7] [8] [9].

In its' simplest form, JMRAM combines superconducting electrodes (S) with a ferromagnetic (F) spin valve to form magnetic barrier Josephson Junctions (MJJs). The write-schemes for JMRAM are similar to conventional field-switched MRAM devices wherein JMRAM utilizes bit- and word-write applied magnetic fields to set the magnetization of the free layer in a magnetic spin valve. Contrary to MRAM, readout is not based on magneto-resistance, but rather on the Josephson Effect in the magnetic junctions. Utilizing well-established physics of S/F/S junctions [10], the junction's equilibrium phase can be toggled between a zero and a π state based on the relative magnetization of the ferromagnetic barrier that stores the data, facilitating readout of the bit by a dc-SQUID *via* the Josephson Effect. Our group has recently demonstrated the ability to build a cryogenic magnetic memory unit cell built in a superconducting integrated circuit, however further efforts for materials optimization must be pursued in order to advance and scale this technology.

For incorporation into a superconducting memory solution, JMRAM optimization requires that magnetic layers exhibit single domain switching with abrupt magnetization reversal and low switching fields, at cryogenic temperatures (e.g. $T$ ~ 4 K). JMRAM requirements are further complicated as materials selection and thicknesses must allow MJJs to carry a sufficient critical current ($I_c$), consistent with low error rate operation at 4 K, and the barrier thickness choice must be in one of two possible junction ground states differing in their equilibrium superconducting phase (0 or π). To date, systems comprised of Ni, $Ni_{80}Fe_{20}$ and $(Ni_xFe_{1-x})_{1-y}M_y$ (where M = a normal metal dopant) have demonstrated the ability to carry a measureable critical current and observable 0- π transition thickness, however, the magnetic switching character of these materials may not be sufficient for incorporation into a final memory solution. [5] [11] [6] [12] [7] To this end, it is necessary to examine low temperature magnetic switching and electrical performance of S/F/S systems with novel ferromagnetic barriers. In this work, we show fundamental magnetic and $I_c$ responses of an S/F/S system with a $Co_{56}Fe_{24}B_{20}$ (CFB) ferromagnetic barrier. To date, CFB has been widely examined as a soft magnetic layer for magnetic tunnel junctions, which utilize a room temperature magnetoresistance–based readout. However, this model system has not been examined at temperatures relevant to a cryogenic memory, nor has it been examined in the context of S/F/S heterostructures. Here, we show magnetic and electrical results of ultra-thin CFB, incorporated into a S/F/S heterostructure. Further, we compare the role of incorporating a simple Cu seed layer against a Nb/Cu multilayered buffer, which provides a smoother starting surface for the ultra-thin film system. [13] While the ultimate objective of these efforts is to achieve a cryogenic memory solution, the work described here probes the fundamental properties of low $T$ magnetics and S/F/S character of a new candidate material for this emerging technology.

## II. Experimental

Magnetic Josephson junctions (MJJs) containing ultra-thin CFB barriers have been fabricated. The ferromagnetic (F), CFB, barriers have been deposited using magnetron sputter deposition in an applied field (~50 Oe) from a single target (>99.9% purity) at ambient temperature with a high purity Ar sputter gas and a base pressure of <3x10$^{-9}$ torr. The films were deposited at a 30-degree angle from the wafer normal, producing a gradient in the F layer thickness of 0.6-1.3 nm across a 150 mm wafer. The wedge films allow for a large number of samples to be deposited at the same time under the same conditions, where the only varying component is the F layer thickness. The CFB films have been sputter-deposited onto two separate surfaces: 1) 150 nm Nb/1 nm Cu and 2) 150 nm Nb/(1 nm Cu/0.5 nm Nb)$_6$/1 nm Cu; both film sets have been capped with 5 nm Cu/50 nm Nb. Here, the incorporation of the (1 nm Cu/0.5 nm Nb)$_6$/1 nm Cu starting surface ultimately reduces the Nb surface roughness and its' deleterious impact on magnetic switching. [13]

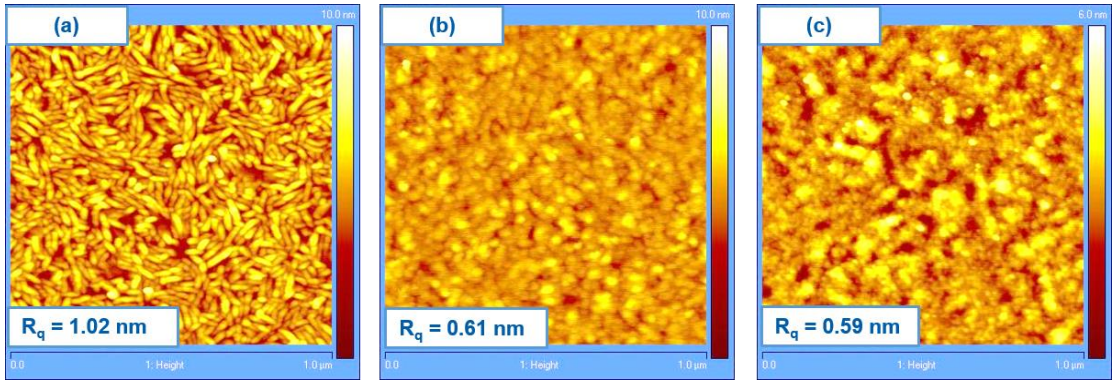

Figure 1: AFM images of (a) 150 nm Nb, (b) 150 nm Nb/5 nm Cu, (c) 150 nm Nb/(1 nm Cu/0.5 nm Nb)$_6$/1 nm Cu

The film thicknesses and amorphous structure of the CFB were confirmed with x-ray reflectivity and diffraction. The thickness gradient of the CFB wedge was determined by measuring the film thickness at multiple locations across the 150 mm wafer and fitting the result to a quadratic function. X-ray diffraction results confirm a (110)-oriented BCC Nb phase and (111)-oriented Cu that may be indexed to an FCC structure. These structural results are not shown here, but are discussed in our previous work. [14]

Tapping-mode atomic force microscopy (AFM) images were collected on the starting surfaces of the films over multiple points within the sample area, prior to CFB deposition. Root mean squared surface roughness ($R_q$) values were obtained within a 1×1μm$^2$ field of view and averaged over three separate measurements.

The measured bulk properties for this specific composition of CFB are as follows. For a 100 nm thick film, the saturation magnetization was measured to be 1625 kA/m with a resistivity (R) of ~ 110 uΩ-cm at $T$ = 10 K. Note, the residual resistivity ratio, RRR = R(300 K)/R(10 K) ~ 1. Using patterned Hall bars and high magnetic field resistance values ($H > 3$ T) to measure the ordinary Hall Effect contribution, the mean free path for both spin channels was determined to be less than 0.1 nm at 10 K, which is expected for an amorphous ferromagnetic metal.

Room temperature measurements of magnetic induction (B) vs. field (H) of continuous, single thickness, CFB films were made using a SHB Looper (Mesa600). The continuous films lithographically patterned into bit arrays comprised of 1×2 μm$^2$ ellipses whose major axis of the shape is parallel to the applied field direction during film growth and the wedge gradient direction. Magnetic characterization of both the blanket films and patterned arrays was carried out at a temperature of $T$ = 10 K, using a vibrating sample magnetometer (Quantum Design, VSM). Magnetic moment (m) vs. H data was collected at a rate of ~1 Oe/second with H applied parallel to the film plane and bit long axis direction; no demagnetization corrections were applied. The resultant hysteresis loops were fit to a standard error function to obtain magnetic coercivity ($H_c$) and switching field distributions (σ) where, σ is defined as the field range at which 68% of the magnetic moments in a film have switched orientation direction.

Separate CFB wedge films were patterned into magnetic Josephson junction (MJJ) devices, comprised of columns of discrete 1×2 μm$^2$ ellipses which spanned the entire 150 mm wafer diameter parallel to the F wedge thickness gradient, to examine the critical current transparency of the F barrier across the deposited thickness range. Here, 5×5 mm$^2$ chips were diced from the wafer and mounted on specially designed non-magnetic 84 pin packages, manufactured by Kyocera, to be inserted into custom liquid He dip probes for cryogenic testing. Measurements of the MJJ I-V relationship was obtained at $T$ = 4 K under nominally zero applied field using a standard kelvin-probe technique. The critical current, $I_c$, was then extracted from a fit to the standard non-hysteretic Josephson junction relationship. These measurements were performed on a selection of chips which comprised a long column parallel to the varying F thickness, providing approximately 350 sample points across the whole F wedge. This high density of samples allows for a more accurate determination of the 0-π phase transition in these films.

## III. RESULTS AND DISCUSSION

The topography (AFM) of the 150 nm Nb/5 nm Cu and 150 nm Nb/(1 nm Cu/0.5 nm Nb)$_6$/1 nm Cu films are shown in Fig. 1. These images display the starting interface on which the magnetic layers were deposited. The starting 150 nm Nb film layer exhibits a small, elongated, rice-grain morphology (Fig. 1(a)) with an rms roughness of $R_q$ = 1.02 nm ± 0.05 nm. Fig. 1(b)-(c) shows the loss of the underlying Nb rice-grain morphology and reduction of the surface roughness to 0.61 nm ± 0.05 nm and 0.59 nm ± 0.05 nm with the incorporation of the 5 nm Cu and (1 nm Cu/0.5 nm Nb)$_6$/1 nm Cu seed layers, respectively.

Fig. 2 shows easy and hard axis switching of continuous 1.5 nm CFB films (measured at $T$ = 300 K) on the two different starting surfaces: 150 nm Nb/5 nm Cu (Fig. 2(a)) and 150 nm Nb/(1 nm Cu/0.5 nm Nb)$_6$/1 nm Cu (Fig. 2(b)). Here, the CFB films exhibit low reversal fields at 300 K with a $M_S$ value ~ 0.38 nWb (1525 kA/m), irrespective of the starting surface for magnetic layer growth. These room temperature results are consistent with conventional MRAM devices that utilize CFB alloys as a soft ferromagnet. Both sets of CFB films exhibit an easy axis $H_c$ < 0.4 mT, squareness (Sq) ~1 and hard axis anisotropy which is induced by the magnetic field applied during the film deposition. Here, the thin (1 nm Cu/0.5 nm Nb)$_6$/1 nm Cu buffer layer at the Nb interface does not significantly improve the CFB switching character at room temperature. This is likely due to the fact that CFB is an amorphous film and as such may not be substantially impacted by seed layers. The strong anisotropy observed in these films is induced during the film growth due to the application of an applied field.

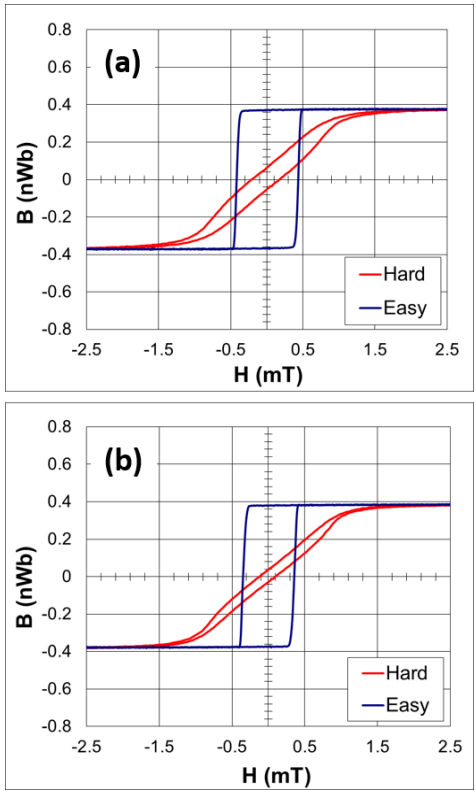

Figure 2: BH Looper plots of magnetic induction vs. applied field of 1.5 nm CFB films deposited on (a) 150 nm Nb/5 nm Cu and (b) 150 nm Nb/(1 nm Cu/0.5 nm Nb)$_6$/1 nm Cu

Fig. 3 shows easy axis hysteresis loops of the patterned 1×2 µm$^2$ elliptical arrays at T = 10 K. Overall, an ~8× expansion in the easy axis $H_c$, and increase in m$_s$ is observed in the patterned films as compared to the room temperature blanket film data (Fig. 2). The increase in the magnetization of ferromagnetic materials when cooled to low temperature is a well-known effect, as related to the Curie-Weiss law, and an increase in $H_c$ with magnetization may be expected. Further, the expansion of the magnetic hysteresis loop and loss of soft magnetic switching character is consistent with low temperature results observed by our group for single layer NiFe films [14] and similar low temperature studies carried out by other groups [15]. While the magnetic switching character of both film sets are similar at room temperature, the CFB films deposited on the 5 nm Cu seed layer exhibit a larger $H_c$, as compared to the film deposited directly on the (1 nm Cu/0.5 nm Nb)$_6$/1 nm Cu starting surface (3.3 mT *vs.* 3.2 mT). The CFB switching distribution on the 5 nm Cu seed exhibits a loss of squareness as demonstrated by the increased σ values (1.2 *vs.* 0.6 mT) as compared to the (1 nm Cu/0.5 nm Nb)$_6$/1 nm Cu buffer. The increased σ values observed in the CFB film deposited on the 5 nm Cu seed may be attributed to variations observed in the surface morphology of this starting surface as compared to the 150 nm/(1 nm Cu/0.5 nm Nb)$_6$/1 nm Cu starting buffer, as suggested from the topography images (Fig. 1).

Fig. 4 shows the thickness dependence of 1×2 µm$^2$ elliptical MJJs with a CFB ferromagnetic barrier, as measured at 4.2 K and zero applied magnetic field. The CFB films deposited on both seed layers exhibit an expected oscillatory decay in the critical current observed described by the following expression:

$$J_C = J_0 * e^{-\frac{d_F}{\xi_{F1}}} * \cos\left(\frac{d_F}{\xi_{F2}} - \varphi\right) \quad (1)$$

where $J_0$ is the maximum critical current density for a device of zero ferromagnetic thickness, $d_F$ is the ferromagnetic thickness, $\xi_{F1}$ is the pair correlation length, $\xi_{F2}$ is the oscillation period, and $\varphi$ is a phase offset. [16] [10] Fit parameters may be extracted from fits to the $I_c$ vs. F thickness data shown in in Fig. 4. It is noted that the absence of multiple minima make the calculated value for the "0 – π" thickness somewhat ambiguous, therefore, only the fitted values for the pair correlation length are reported here.

The pronounced dips in $I_c$ of both wedge films, regardless of the buffer layer materials, at 0.8 nm shows the characteristic of a change in the equilibrium Josephson phase of the S/F/S structure (i.e. the "0 – π" thickness) and appears related to the CFB material intrinsically. This indicates that the differing buffer layers do not affect the magnetic properties of the F layer in the sample. It can be seen, both from graphical trends and the extracted fitting parameters, that the wedge film with the (1 nm Cu/0.5 nm Nb)$_6$/1 nm Cu seed layer buffer shows an overall lower critical current carrying capacity with respect to the single Cu buffer layer sample. It also appears to show a slightly slower decay rate of $I_c$ across the CFB thickness range tested here. This suggests that the incorporation of additional layers and interfaces into the junction stack has a more complex impact on the overall current transparency of the MJJ.

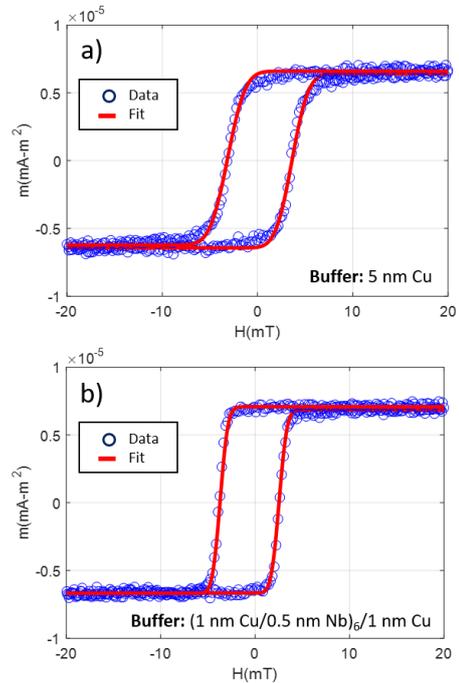

Figure 3: T = 10 K VSM data of patterned 1 x 2 um elliptical arrays of 1.5 nm CFB films deposited on (a) 150 nm Nb/5 nm Cu and (b) 150 nm Nb/(1 nm Cu/0.5 nm Nb)$_6$/1 nm Cu. Raw data and fits are both shown here.

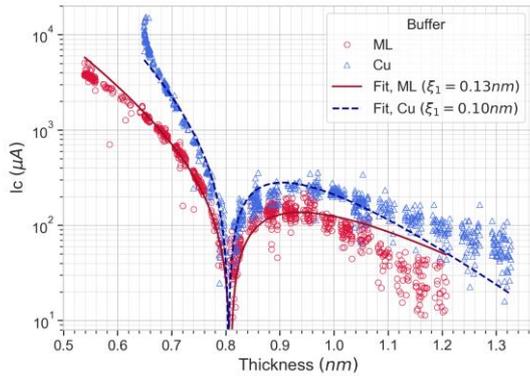

Figure 4: Critical current ($I_c$) vs thickness trend across CFB wedges deposited on 150 nm Nb/5 nm Cu and 150 nm Nb/(1 nm Cu/0.5 nm Nb)$_6$/1 nm Cu buffer layers. The extracted pair-correlation lengths from fits to Eq. 1 are shown as lines and presented in the legend. ML corresponds to the multilayed Cu/Nb buffer layer.

Data in the thicker portion of the F wedge range shows increased scatter with respect to the thinner and middle thickness ranges. This increased scatter is believed to be predominantly due to the non-initialized magnetic state of the MJJ, which provides a random "fraunhofer-shift" in the junction critical current that cannot be easily predicted or decoupled when performing the measurements in 0-applied field conditions. [16] [5] [9] Fraunhofer switching measurements of these devices in the thicker regions of the wedge film were not performed here due to the limitations of the testing setup. In the future, performing these types of measurements may provide higher fidelity data in this region of the sample.

Divergence of the theoretical curve from the data can be seen near the edges of the thickness range as well. It is suspected that this is due to the "fraunhofer-shift" effect described above or unintended shadowing effects of the oblique deposition technique near the edges of the wafer. Future measurement of films with overlapping, but different, thickness ranges may provide insight into the exact origins of this deviation from theoretical behavior.

IV. CONCLUSIONS

Here, we have shown magnetic and electrical properties of ultra-thin CFB films. At room temperature, these films show near ideal soft magnetic switching, making this system of interest for a free layer in memory applications which utilize a magnetic spin valve. When patterned into arrays and measured at low temperature, this system shows an expansion of the magnetic hysteresis loop. Specifically, the incorporation of the (1 nm Cu/0.5 nm Nb)$_6$/1 nm Cu buffer shows reduced $H_c$ and σ of the magnetic hysteresis of patterned arrays, measured at 10 K, relative to the 5 nm Cu seed layer. Electrical data of fabricated MJJs shows a dip in the $I_c$ at 0.8 nm suggesting a 0-π phase shift, at this thckness. A slight reduction in the MJJ $I_c$ is observed with the incorporation of the (1 nm Cu/0.5 nm Nb)$_6$/1 nm Cu buffer which suggests that incorporating additional layers into S/F/S structures may lead to a reduction in $I_c$ transparency, however the 0-π transition thickness remains unchanged in the case of amorphous CFB.